\documentclass[]{revtex4}
\usepackage{hyperref}
\begin{document}

\title{Super-group field cosmology in Batalin-Vilkovisky  formulation}

 \author{ Sudhaker Upadhyay}
  \email{ sudhakerupadhyay@gmail.com}
\affiliation {Centre for Theoretical Studies,
Indian Institute of Technology Kharagpur, Kharagpur, 721 302, WB, India}

\begin{abstract}
 In this paper, we study the  third quantized super-group field cosmology, a model in 
 multiverse
 scenario, in Batalin-Vilkovisky  (BV) formulation.  Further, we propose the superfield/super-
 antifield dependent  BRST symmetry transformations. Within this formulation, we establish 
 connection between the two
 different solutions of the quantum master equation  within the BV formulation.
  \end{abstract}

\maketitle 
\section{Introduction}
The construction of a consistent theory of quantum gravity continues to be one of the
major open problems in fundamental physics. Such a theory is very essential for the understanding of fundamental
issues such as the origin of the Universe, the final evaporation of black holes,
and the structure of space and time.  Several approaches to quantum gravity have been developed, with a remarkable convergence \cite{oriti}. The loop quantum gravity, a
 background-independent approach,   is one of the powerful candidates quantizing gravity in mathematically
rigorous and in non-perturbative way \cite{rov,thi}.  The development started with the introduction of  Ashtekar-Barbero  variables,  the densitized triad and the Ashtekar connection  \cite{as0, as, as1,as2,li,ge}. However,
Loop quantum cosmology  is the result of applying principles of loop quantum gravity  to cosmological settings.  The ensuing framework
of loop quantum cosmology was introduced by Martin Bojowald  \cite{boj}. The mathematical structure
of loop quantum cosmology is presented in Ref. \cite{ashte}. Loop quantum cosmology is
constructed via a truncation of the classical phase space of general relativity to spatially
homogeneous situations, which is then quantized by using the methods and results
of loop quantum gravity. The quantization
of geometric operators are thereby transferred to the truncated models.

However, the group field theories have been proposed
as a kind of second quantization of canonical loop quantum
gravity, in the sense that its
canonical wave function  turns into a dynamical (quantized) field \cite{fre, oriti1}. 
 The group field theories are basically described by the
field theories on group manifolds (or their Lie algebras) which  provide a
 background-independent
third quantized formalism for  gravity in any dimensions and signature \cite{ori1,ori2}. 
In the group field theories, both the geometry and the topology are  dynamical.
The Feynman diagrams of such theories have
an interpretation of the  spacetimes and  therefore the quantum amplitudes for these diagrams 
can be interpreted as   algebraic realization of a path
integral description of gravity  \cite{ori,per}.

  The topology changing processes can not be analyzed by second quantization approach. This brings group field theory into the conceptual
framework of ``third quantization",   for a
rather appealing idea  \cite{bu,pi,pe,ma, ic,kim,faizal0,faizal1}.  The third quantization is a field theory on the space of geometries, rather than
spacetime, which also allows for a dynamical description of topology change \cite{str}.
  Remarkably, the third quantization of loop quantum gravity leads to the group field theory \cite{ab, sm, ab1, ta}. The Wheeler-De Witt (mini-superspace) approximations of the
group field theory results in the group field cosmology  
\cite{aa, qi, gu, ew, qi1, cal, gl,sg}.

On the other hand, supersymmetry is an attractive concept whose basic feature is a transformation which 
relates bosons to fermions and vice-versa \cite{van, free}. One of its more significant features is that the presence
of local supersymmetry naturally entails spacetime to be curved. The promotion of
 supersymmetric theory  to a gauge symmetry has resulted in  supergravity \cite{van}. At large scales, supergravity permits to make the same predictions for classical tests as general relativity.
 The supersymmetry has also been testified as a prominent candidate for the dark matter \cite{sa}. 
The supersymmetrization of group field cosmology has  been made recently which is known as super-group field cosmology 
 \cite{fai0,sud}. Remarkably, the   super-group field cosmology turns out to be gauge invariant and according to standard quantization principles  a gauge symmetric models must be quantized after fixing a gauge as they possesses  some spurious  degrees of freedom. Hence to get rid of spurious degrees of freedom  one should fix the gauge in case of super-group field cosmology. 
Such gauge-fixing adds the Faddeev-Popov ghosts in the void  functional (the vacuum
 functional  of third quantization) of the theory which helps in defining the physical Hilbert state of the
 effective theory.  The fermionic rigid 
 BRST transformation   and thus unitarity of super-group field cosmology has been studied recently  \cite{fai}.
The Slavnov-Taylor identity and renormalizability of the theory 
has also been proved \cite{sud}.  The cosmological Slavnov-Taylor  identity of the Einstein-Hilbert action coupled to a single inflation field is  derived recently  in the Arnowitt-Deser-Misner   formalism \cite{qu}. BRST algebra for the mixed Weyl-diffeomorphism residual symmetry
is derived in Ref. \cite{fran}. The BRST symmetry gets relevance  in many more contexts also
\cite{faiz0}. 

 The  BV  formalism \cite{henna,wein,bv1,bv2,bv3}, also known as the field/antifield and the BRST-BV 
 approaches, is one of the most powerful techniques in the study gauge field theories 
which  deals with very general gauge 
theories, including those with open or reducible gauge symmetry algebras.
The BV method of quantization provides a convenient way of analysing the possible violations of 
symmetries  \cite{wein}.  Basically, it is  used  to perform the gauge-fixing
in quantum field theory, however, it was also applied to other problems like analysing
possible deformations of the action and anomalies.
 The  BV approach is a successful for studying the 
manifestly Lorentz invariant formulation of the string theory \cite{sei}.
Utilizing  variational tricomplex, a covariant procedure  for deriving the classical BRST charge   from a given BV master action is proposed recently \cite{shar}.
Using BRST-BV formulation of relativistic dynamics, arbitrary spin massless and massive field propagating in flat space and arbitrary spin massless fields propagating in AdS space are 
studied \cite{met}. The BV formalism for the  theory of super-group field cosmology
has not studied yet. We try to write the quantum master equation for extended quantum action.  The quantum master equation gives relation between different 
sets of greens functions and vertices. 

 The  finite field-dependent BRST (FFBRST) transformation  has been investigated  originally in \cite{sdj}. 
Further it has been found many implications in the diverse gauge  theories 
\cite{sb,sdj,sdj1, smm,fs,sudpr,srb, susk,ssb,sudd,sud0,sud001,sb000,sud21,sud31,sud01,rbs,rs,fsm,sud33,srm}. 
For example, more recently, 
 the gauge-fixing and ghost terms corresponding to Landau and maximal Abelian gauge have  been   
 rendered for the  
  Cho-Faddeev-Niemi  decomposed $SU(2)$ theory  using FFBRST transformation
  \cite{sud0}. However, the connection between linear and non-linear gauges for perturbative quantum gravity  at both classical and quantum level has been established through FFBRST formulation \cite{sud001,sb000}.
  The  quantum gauge freedom studied by gaugeon formalism has also been  addressed
  for quantum gravity \cite{sud21} as well as for Higgs model \cite{sud31} utilizing FFBRST technique.
The FFBRST transformations get relevance for the lattice gauge theory \cite{rs} and the
relativistic point particle model \cite{rbs}.  
With the help of FFBRST transformations it has also been proved that   the problems associated with Virasoro constraints
in worldsheet gravity are the gauge artifact \cite{sud33}.
Recently, Regge--Teitelboim cosmological model is quantized  from the FFBRST point of view \cite{sbp}.
The FFBRST  symmetries has also been derived in Friedmann-Robertson-Walker cosmological models \cite{supa}. Moshin and Reshetnyak in Ref. \cite{ale}  systematically incorporates the case  of BRST-antiBRST symmetry in Yang-Mills theories
 within the context of finite transformations that deals
 with the case of a quadratic   dependence  on the corresponding parameters.
 Further,   the concept of finite BRST-antiBRST
 symmetry to the case of general gauge theories has been extended in Refs. \cite{ale1, mos1},
 whereas Ref. \cite{mos} by the same authors
 generalizes the corresponding parameters
 to the case of arbitrary  Grassmann-odd
 field-dependent parameters, as compared
 to the so-called ``potential" form of parameters
\cite{ale,ale1,mos1}.

The BV  formulation and its connection to superfield/super-antifield dependent BRST
symmetry have not been discussed so-far
for the third quantized super-group field cosmology. 
This provides us a glaring omission. Here we remark that 
the FFBRST transformation turns to superfield/super-antifield dependent BRST transformation
as the super-group field cosmology exists for super-manifold. 

In the present paper we quantize the super-group field cosmology using BV-BRST method.
To do so, we introduce the   different gauge-fixing conditions and corresponding 
super-ghosts in the theory. We further demonstrate the infinitesimal BRST symmetry
for the super-group field cosmology. This model of multiverse is also analysed 
through BV formulation where we extend the configuration space by introducing super-antifields for
each set of superfields. The extended quantum actions corresponding to the   different gauge-fixing conditions are shown the different solutions of the quantum master equation.
Further we generalize the BRST symmetry by making the parameter of transformation 
superfield/super-antifield dependent. Remarkably, we found that
such generalized BRST transformation leads to the non-trivial Jacobian for the
path integral measure. We explicitly compute the Jacobian for superfield/super-antifield dependent
BRST transformation at general level. Furthermore, we observe that
for specific choice of superfield/super-antifield dependent transformation parameter 
the Jacobian leads  the theory from one gauge to another.

The   paper is presented as follows. In section II, we discuss the  supersymmetric
group field cosmology in various gauges. The  supersymmetric
group field cosmology in BV formulation is analysed in the section III. The
discussion on superfield/super-antifield dependent BRST symmetry  is reported in section IV.  
In the last section we made a concluding remarks. 
\section{Super-group field cosmology and their BRST invariance}
Let us start by recapitulating the progress made in \cite{sud} for the loop quantum cosmology of the spatially flat, homogeneous and isotropic universe having massless scalar field  as the bulk.
The loop quantum gravity is described as a gauge theory where the dynamical variables are the 
Ashtekar-Barbero connection $K^i_a$ and canonical momenta, the densitized triad $E_i^a$. Here, $a, b = 1, 2, 3$ are the usual space index
(referring to the tangent space $T_x(\Sigma)$ at $x$). 
 These variables are defined in terms of co-triads $e^i_a$ as $K_a^i(x)=K_{ab}(x)e^{bi}(x)$ and $E_i^a =|\mbox{det} (e^i_a)|e^a_i(x)$, where   the extrinsic curvature $K_{ab}$  in terms of lapse
 $N$ and shift $N_a$ can be written as
\begin{eqnarray}
K_{ab}=\frac{1}{2N}\left(\dot{h}_{ab}-\nabla_aN_b-\nabla_bN_a \right),
\end{eqnarray}
a covariant derivative on (mini-)superspace of geometries $\nabla_a$. The four dimensional metric in this case is described by a three metric $h_{ab}$ given as 
\begin{eqnarray}
h^{ab}=\delta^{ij}e^a_ie^b_j=e^a_ie^b_i,
\end{eqnarray}
where triad $e^a_i$ is the inverse of co-triad $e^i_a$.

We  construct 
 the classical action for the super-group field cosmology, given by \cite{fai0}
 \begin{eqnarray}
S_{0} =\sum_\nu \int d\phi \, \,    \left[D^2 \{\Omega_i^{\dagger} (\varrho) \nabla_a^2 \Omega^i
 (\varrho) 
  + \omega_i^a (\varrho) \omega^i_a(\varrho)  \}\right]_|,
\end{eqnarray}
where $``|"$ stands for Grassmann variable which describes a space with with supersymmetric degrees of freedom at $\theta_a = 0$ and (mini-)superspace variables $(\nu, \phi, \theta) :=\varrho$.  Here
$\Omega(\nu, \phi, \theta) $ and $ \Omega^{\dagger}  (\nu, \phi, \theta)$ are two complex scalar super-group fields.
The super-covariant derivatives of $\Omega^i(\varrho)$ and $\Omega^{i \dagger}(\varrho)$ are defined by
\cite{fai0}
\begin{eqnarray}
  \nabla_a  \Omega^i(\varrho)&=& D_a\Omega^i(\varrho) -i f_{kj}^i\Gamma^k_a(\varrho) \Omega^j(\varrho),\nonumber\\
\nabla_a \Omega^{i \dagger}(\varrho)  &=& D_a \Omega^{i \dagger}(\varrho)  
+ i  f_{kj}^i\Omega^{k \dagger}(\varrho)  \Gamma^j_a (\varrho), 
\end{eqnarray}
where super-derivative $ D_a = \partial_a + K^b_a \theta_b$.
The field-strength for  a matrix valued spinor field ($ \Gamma^i_a $) is given by
$
 \omega^i_a (\varrho) = \nabla^b \nabla_a \Gamma^i_b(\varrho) $.
It is evident that this classical action is invariant  under the following gauge transformations:
\begin{eqnarray}
  \delta \Omega^i(\varrho) &=&  if_{kj}^i\Lambda^k (\varrho)\Omega^j(\varrho) ,\nonumber\\
\delta \Omega^{i \dagger}(\varrho)  &=& -i f^i_{kj}\Omega^{k\dagger}(\varrho) \Lambda^j(\varrho), \nonumber\\
 \delta \Gamma^i_a(\varrho) &=& \nabla_a \Lambda^i(\varrho),
\end{eqnarray}
where the bosonic transformation parameter $\Lambda^i$ is   infinitesimal in nature. 
 
In the path integral formulation, due to this gauge symmetry there exist infinitely many 
$^{(\Lambda)}\Gamma_a^i$ that are physically equivalent to $\Gamma_a^i$. This produces divergences 
in the functional integral. To quantize this theory consistently it is necessary to eliminate redundant 
gauge degrees of freedom by choosing a particular gauge. Here we choose a general gauge condition for this theory as
\begin{equation}
{\cal G}^i[ \Gamma^i_a (\varrho) ] =0.\label{gau}
\end{equation}
The above gauge condition can be incorporated in the theory at quantum level by adding the
corresponding gauge-fixed action to the classical action. According to the Faddeev-Popov procedure the 
 gauge-fixing condition leads to the ghost term in the effective theory.
The linearised gauge-fixing action  corresponding to the gauge (\ref{gau}) together with the induced ghost term is given by
\begin{equation}
 S_{gf+gh} = \sum_\nu \int d\phi \, \,   D^2\left[  B_i (\varrho){\cal G}^i[ \Gamma^i_a (\varrho) ] +
 \bar c_i(\varrho)s_b {\cal G}^i[ \Gamma^i_a (\varrho) ] \right]_|,\label{gf}
\end{equation}
where  $  B^i( \nu, \phi, \theta) $ is the
Nakanishi-Lautrup auxiliary superfield, $   {c}^i(\varrho)  $
and   $ \bar{c}^i (\varrho) $ are   
the ghost and anti-ghost superfields respectively, and, 
$s_b$ denotes the Slavnov variation.
Now, the total effective action for super-group field cosmology for general gauge choice reads
$
S_T =  S_{0}+  S_{gh}+  S_{gf}. 
$

Now we consider an specific choice of Landau type gauge 
  ${\cal G}^i=D^a\Gamma^i_a(\varrho) =0$, then
the total action $S_T$ gets following identification  \cite{fai}:
\begin{eqnarray}
S_T =  S_{0}+  \sum_\nu \int d\phi \, \,   \left[D^2\{ B_i (\varrho)D^a\Gamma^i_a(\varrho)+ \bar c_i(\varrho) D^a\nabla_a c^i (\varrho) ] \}\right]_|.\label{ori}
\end{eqnarray} 
Furthermore, to analyse the theory in  massless Curci-Ferrari type gauge (a non-linear 
gauge) we perform the following shift in auxiliary superfield:
$B^i( \nu, \phi, \theta) \longrightarrow B^i(\varrho)  -\frac{1}{2}f^i_{jk} \bar c^j (\varrho) 
c^k(\varrho)$. Performing such shift
the total effective action corresponds to a non-linear gauge  as follows
\begin{eqnarray} 
S_T &= &S_0+ \sum_\nu \int d\phi \, \,   
 \left[D^2\left\lbrace B_i (\varrho) D^a \Gamma^i_a (\varrho)+  \bar c_i(\varrho) D^a \nabla_a c^i
(\varrho) \right.\right.\nonumber\\
&-&\left.\left. \frac{1}{2}f_i^{jk} B^i(\varrho)\bar c_j(\varrho)c_k(\varrho) -\frac{1}{8} f_{jk}^if_{i}^{lm}\bar c^j (\varrho) \bar c^k(\varrho)c_l(\varrho) c_m(\varrho)\right\rbrace\right]_|.\label{acta}
\end{eqnarray} 
These effective actions (\ref{ori}) and (\ref{acta}) are invariant under the
following   (third quantized) infinitesimal BRST transformations \cite{fai} 
\begin{eqnarray}
  \delta_b\,\Omega^i(\varrho)&= &if_{kj}   ^ic^k (\varrho)\Omega^j(\varrho) \ \delta\lambda, \nonumber \\
\delta_b\,  \Omega^{i \dagger}(\varrho)  &= & -i f^i_{kj}\Omega^{ \dagger k}(\varrho)c^j(\varrho)\ \delta\lambda, 
\nonumber \\ 
\delta_b\, c^i( \varrho)&= &\frac{1}{2} f^i_{kj}c^{  k}(\varrho) c^j(\varrho)\ \delta\lambda, 
\nonumber \\
\delta_b\, \Gamma^i_a (\varrho) &= & \nabla_a c^i(\varrho)\ \delta\lambda, 
\nonumber \\ 
\delta_b\, \bar{ c} ^i(\varrho) &= & B^i(\varrho)\ \delta\lambda,
 \nonumber \\
\delta_b\, B^i(\varrho) &= &0,\label{brs}
\end{eqnarray}
where $  \delta\lambda$ is an infinitesimal, space-time independent anticommuting  parameter. 
It is easy to verify that the above transformations are nilpotent of order two,  i.e.,  $\delta_b^2 =0$. 
With  the help of above BRST transformation, one can write the sum of gauge-fixing and ghost parts of the action given in (\ref{gf}) 
as a BRST variation of gauge-fixed fermion  $\Psi=D^2 \{\bar{c}_i  (\varrho) {\cal G}^i[ \Gamma^i_a (\varrho) ] \}$  as follows
\begin{eqnarray}
S_{gf+gh} = \sum_\nu\int d\phi \ s_b \Psi.
\end{eqnarray}
 Such analysis will be helpful to establish theory in BV formulation.
 \section{Super-group field cosmology in Batalin-Vilkovisky formulation}
 To establish the theory in BV formulation we need to introduce super-antifield  corresponding to each superfield having opposite statistics. In terms of superfield/super-antifield, the generating functional for the super-group field cosmology  in Landau type gauge is,
\begin{eqnarray}
Z_L[0]  &=&\int {\cal D} M\ e^{-W_{L}[ \Phi , \Phi^\star]}\nonumber\\
&=&\int {\cal D} M \exp\left[- \left(S_0+  \sum_\nu \int d\phi \, \, \left[\Gamma^{a\star}_{1i} \nabla_a c^i+c_{1i}^\star f^i_{kj}c^kc^j+\bar{c}_{1i}^\star B^i \right]_| \right) \right],
\label{zlin}
\end{eqnarray}
where $W_L$ is the extended quantum action. 
The gauge-fixed fermion for the super-group field cosmology  in Landau type gauge has the following expression:
\begin{eqnarray}
\Psi_L = D^2 [\bar{c}_i (\varrho) D^a\Gamma^i_a (\varrho) ]. 
\end{eqnarray}
Now we compute the following super-antifields  for corresponding to Landau gauge  
\begin{eqnarray} 
\Omega^{i\star}_1 &=& \frac{\delta \Psi_L}{\delta \Omega_i} = 0, \, \, \, \, \, \, 
\Omega^{i\dagger\star}_1 = \frac{\delta \Psi_L}{\delta \Omega^{\dagger}_i} = 0,   \nonumber \\
  c_{1i}^\star  &=&   \frac{\delta \Psi_L}{\delta c^i} = 0, \, \, \, \, \, \,
\bar{c}_{1i}^\star =  \frac{\delta \Psi_L}{\delta \bar{c}^i} = D^2D_a\Gamma^a_i, \nonumber \\
\Gamma^{a\star}_{1i}  &=& \frac{\delta \Psi_L}{\delta \Gamma_a^i} =  -D^2D^a\bar c_i.
\label{antil}
\end{eqnarray}
However, the generating functional for the super-group field cosmology  in the non-linear gauge in terms of 
superfields/super-antifields is given by,
\begin{eqnarray}
Z_{NL}[0]  &=&\int {\cal D} M\ e^{-W_{NL}[ \Phi , \Phi^\star, \tilde\Phi, \tilde\Phi^\star]}\nonumber\\
&=&\int {\cal D} M 
e^{ - \left(S_0+  \sum_\nu \int d\phi \, \, \left[ \Gamma^{a\star}_{2i}\nabla_a c^i +c_{2i}^\star
\left(\frac{1}{2} f^i_{kj}    c^{  k}  c^j \right) +\bar{c}_{2i}^\star  B^i  
\right]_|\right)  }.
\label{znil}
\end{eqnarray}
The expression for the gauge-fixing fermion for the non-linear gauge is given by
\begin{eqnarray} 
\Psi_{NL}  =  D^2 \left[\bar{c}_i  (\varrho)\left( D^a\Gamma^i_a (\varrho) -\frac{1}{4}f^{ijk} \bar c_j 
(\varrho) c_k(\varrho) \right) \right].
\end{eqnarray}
The super-antifields corresponding to above gauge-fixing fermion are identified by
\begin{eqnarray} 
\Omega^{i\star}_2 &=& \frac{\delta \Psi_{NL}}{\delta \Omega_i} = 0, \, \, \, \, \, \, 
\Omega^{i\dagger\star}_2 = \frac{\delta \Psi_{NL}}{\delta \Omega^{\dagger}_i} = 0,   \nonumber \\
  c_{2i}^\star  &=&   \frac{\delta \Psi_{NL}}{\delta c^i} = -D^2\left(\frac{1}{4}f_i^{jk}\bar c_j \bar 
  c_k\right),\nonumber \\
\bar{c}_{2i}^\star &=&   \frac{\delta \Psi_{NL}}{\delta \bar{c}^i} =  D^2\left[ D_a\Gamma_i^a -\frac{1}
{2}f_i^{jk}
\bar c_j c_k\right], \nonumber \\
\Gamma^{a\star}_{2i}  &=& \frac{\delta \Psi_{NL}}{\delta \Gamma_a^i} =  -D^2D^a\bar c_i.  
\label{antinl}
\end{eqnarray}
The difference between the non-linear and linear extended quantum actions are given by
\begin{eqnarray} 
 W_{NL} -  W_{L}  &=&    \sum_\nu \int d\phi \, \, \left[  c_{2i}^\star(\varrho)
\left(\frac{1}{2} f^i_{kj}    c^{  k} (\varrho) c^j (\varrho)\right)  +(\bar{c}_{2i}^\star(\varrho)-\bar{c}_{1i}^\star(\varrho))B^i(\varrho)
\right]_|.
\label{diff}
\end{eqnarray}
Here we note that these extended quantum actions, $W_{\Psi}[\Phi,\Phi^\star]\equiv (W_{NL}, W_{L})$, are 
solutions of the following mathematically rich
relation so-called  quantum master equation,  
\begin{equation}
\Delta e^{iW_{\Psi }[\Phi,\Phi^\star] } =0,\ \
 \Delta\equiv (-1)^{\epsilon}\frac{\partial_l}{
\partial\Phi}\frac{\partial_l}{\partial\Phi^\star}.
\label{mq}
\end{equation}
 
 In the next section, we shall establish a map between the two generating functionals
corresponding to the above extended actions using the technique of superfield/super-antifield dependent BRST transformations.
\section{Generalized BRST symmetry for super-group field cosmology }
 In this section, we   analyse the  superfield/super-antifield   dependent BRST transformation which is characterized by the superfield/super-antifield   dependent BRST  parameter.
For this purpose, we  first write the usual BRST transformation 
 given in (\ref{brs})  in compact form as following:
 \begin{eqnarray}
\Phi_\alpha'(x)-\Phi_\alpha(x)&=&\delta_b  \Phi_\alpha(x)= s_b  \Phi_\alpha(x)\delta\lambda ={\cal R}_\alpha(x) \delta\lambda, 
 \end{eqnarray}
where ${\cal R}_\alpha(x)(s_b  \Phi_\alpha(x))$   is the  Slavnov variations of the collective superfield $\Phi_\alpha(x)$ 
satisfying $\delta_b {\cal R}_\alpha(x) =0$.
Here the infinitesimal transformation  parameter $\delta\lambda$  is  a Grassmann parameter 
and doesn't depend on any superfield/super-antifield.

Now, we propose the  superfield/super-antifield dependent BRST transformation   as follows
 \begin{eqnarray}
\delta_b  \Phi_\alpha(x)&=&\Phi_\alpha'(x)-\Phi_\alpha(x)={\cal R}_\alpha(x) \Lambda [\Phi,\Phi^\star],\label{qg}
 \end{eqnarray}
 where the Grassmann parameter  $\delta\lambda$ is replaced by
  $\Lambda [\Phi,\Phi^\star]$ which depends  on the superfield/super-antifield explicitly.  
  The novelty of  superfield/super-antifield dependent BRST transformation is that   
   although being symmetry of the extended action such  transformation does not leave the
   functional measure invariant and leads a non-trivial local Jacobian.  
      
      Now,  we evaluate the Jacobian of functional measure
      under superfield/super-antifield dependent BRST transformation (within functional integral) as follows
 \begin{eqnarray} 
  Z'_{L}  [0] &=& \int {\cal D}M ( \text{sDet} J[ \Phi , \Phi^\star] )\exp\{- W_{L}[ \Phi , \Phi^\star]\}, \nonumber \\
 &=& \int {\cal D} M e^{- (W_{L}[ \Phi , \Phi^\star]-i \text{sTr} \ln J[\Phi, \Phi^\star])},
 \label{deltaZ}
 \end{eqnarray}
 where  $ Z'_{L}$ denotes the generating functional under change of variables.  
The Jacobian matrix for the superfield/super-antifield  dependent BRST transformation is computed by
 \begin{eqnarray}
 J_\alpha^{\, \, \beta} [ \Phi , \Phi^\star] = \frac{ (\delta \Phi'_\alpha, \delta \tilde\Phi'_\alpha)}{(\delta \Phi_\beta, \delta \tilde\Phi_\beta)} &=& \delta_\alpha^{\, \, \beta}  + \frac{\delta {\cal R}_\alpha(x) }{\delta \Phi_\beta} \Lambda [\Phi, \Phi^\star] + {\cal R}_\alpha(x) \frac{\delta \Lambda [ \Phi, \Phi^\star]}{\delta \Phi_\beta}.
 \label{det} 
 \end{eqnarray} 
The nilpotency property of the BRST transformation (i.e. $s_b^2 = 0$) and relation (\ref{det})  yield 
 \begin{eqnarray} 
 \text{sTr} \ln J[ \Phi , \Phi^\star] = -\ln ( 1+ s_b \Lambda [\Phi, \Phi^\star ] ),
 \label{jaco}
\end{eqnarray} 
which simplifies to
\begin{eqnarray} 
\text{sDet} J[ \Phi , \Phi^\star] = \frac{1}{1  + s_b \Lambda[\Phi, \Phi^\star]}.
\end{eqnarray}
With this Jacobian the relation (\ref{deltaZ}) modified by
\begin{eqnarray} 
 Z'_{L}  [0]  = \int {\cal D} \Phi \exp   \bigg(-W_{L}[ \Phi , \Phi^\star]- i\ln ( 1+ s_b \Lambda [\Phi, \Phi^\star ])\bigg).
 \label{deltabZ}
 \end{eqnarray}
 This means that the effective quantum action gets change  under the superfield/super-antifield dependent BRST transformation characterized by an arbitrary $\Lambda[\Phi,\Phi^\star]$.
Now we compute the specific change in the effective action of the super-group field cosmology under the superfield/super-antifield dependent BRST transformation having an specific $\Lambda[\Phi,\Phi^\star]$.
Therefore, we construct the specific parameter as follows
\begin{eqnarray}
\Lambda[\Phi, \Phi^\star] &=& \sum_\nu \int d\phi \ \bar c^l B_l(B^2)^{-1} \bigg( \exp \bigg\{ -i \left[  c_{2i}^\star(\varrho)
\left(\frac{1}{2} f^i_{kj}    c^{  k} (\varrho) c^j (\varrho)\right) \right. \nonumber\\ 
&+& \left. (\bar{c}_{2i}^\star(\varrho)-\bar{c}_{1i}^\star(\varrho))B^i(\varrho)
\right] \bigg\} -1 \bigg)_|,
\label{res1}
\end{eqnarray}
where $B^2=:B^iB_i$.
Now the Slavnov variation gives 
 \begin{eqnarray}
s_b\Lambda[\Phi, \Phi^\star] &=& \sum_\nu \int d\phi \  \exp  \bigg(  -i \left[  c_{2i}^\star(\varrho)
\left(\frac{1}{2} f^i_{kj}    c^{  k} (\varrho) c^j (\varrho)\right) \right. \nonumber\\ 
&+& \left. (\bar{c}_{2i}^\star(\varrho)-\bar{c}_{1i}^\star(\varrho))B^i(\varrho)
\right]  \bigg)_|-1 .
\end{eqnarray}
 The Jacobian (\ref{jaco}) for the  parameter (\ref{res1}) gets the following identification:
 \begin{eqnarray} 
 i \ln ( 1 + s_b \Lambda [\Phi, \Phi^\star ]) &=& \sum_\nu \int d\phi \left[  c_{2i}^\star(\varrho)
\left(\frac{1}{2} f^i_{kj}    c^{ k} (\varrho) c^j (\varrho)\right) 
+(\bar{c}_{2i}^\star(\varrho)-\bar{c}_{1i}^\star(\varrho))B^i(\varrho)
\right]_|. 
 \label{sb}
 \end{eqnarray} 
Therefore,  from the expressions (\ref{diff}),  (\ref{deltabZ}) and (\ref{sb}) it is evident that 
\begin{eqnarray} 
  Z'_L [0] = Z_{NL}[0].
 \end{eqnarray}
 Hence, it is observed that the superfield/super-antifield dependent BRST transformation with   parameter  (\ref{res1}) correlate two different solutions of the quantum master equation.
\section{Concluding remarks}
In this paper we consider  the   supersymmetric group field cosmology, a model for homogeneous and isotropic  multiverse.  In the multiverse scenario, the gauge and the matter sectors describe the
different universes.  
We study the supersymmetrization of group field cosmology which is a gauge invariant theory.
The third quantized infinitesimal BRST transformations
for the super-group field cosmology is demonstrated  for the
Landau type and Curci-Ferrari type gauge-fixing conditions. 
Further, the super-group field cosmology is studied in the context of BV formulation. This may provide a consistent quantum description of supersymmetric group field cosmology. 
In this approach we introduce the super-antifield corresponding to each superfield of the theory.
The extended actions of the super-group field cosmology are shown the solutions of the 
mathematically rich quantum master equation.  The quantum master equation
is an important identities for such model. 

Furthermore, we have generalized the BRST transformation   of the
 super-group field cosmology by making the transformation parameter superfield/super-antifield dependent.
We have found that under
the infinitesimal BRST transformation  both the effective action and  the
 functional measure remain
invariant. However, under
 superfield/super-antifield dependent 
 BRST transformation  only effective action remains invariant while
  the functional measure does not. We have computed the Jacobian for functional measure under 
arbitrary superfield/super-antifield dependent BRST 
transformation explicitly.   Remarkably, we have found that under specific  superfield/super-antifield dependent BRST 
transformation the Jacobian switches the 
void functional from Landau type gauge-fixing condition to Curci-Ferrari type gauge-fixing condition.
Such analysis may be an important  towards the
establishment of the quantum theory of super-group field cosmology.
Further implications of  superfield/super-antifield dependent BRST 
transformation  on the multiverse model, for example calculating the certain
observables, will be subject of interest.

\end{document}